\documentclass[12pt]{article}
\usepackage{amsmath}
\usepackage{graphicx,psfrag,epsf}
\usepackage{enumerate}
\usepackage[round]{natbib}

\usepackage{graphicx}
\usepackage{amsmath}
\usepackage{amssymb}
\usepackage{listings}
\usepackage{paralist}
\usepackage{booktabs}
\usepackage{color}
\usepackage{subcaption}
\usepackage[]{setspace}
\usepackage{xcolor}

\newtheorem{theorem}{Theorem}
\newtheorem{proof}{Proof}
\newtheorem{corollary}{Corollary}

\usepackage{times}
\usepackage{inconsolata}

\usepackage{bm}
\usepackage{natbib}

\usepackage{authblk}            % Multiple authors

\usepackage[plain,noend]{algorithm2e}

\usepackage{url} % not crucial - just used below for the URL 

\newcommand{\blind}{0}

\definecolor{dodgerblue}{HTML}{005A9C}
\definecolor{burntorange}{HTML}{BF5700}
\definecolor{forestgreen}{HTML}{228B22}

\newcommand{\response}[1]{{\color{black} #1}}

\newcommand{\dee}{\mathrm{d}}
\DeclareMathOperator{\pr}{Pr}

\begin{document}

\def\spacingset#1{\renewcommand{\baselinestretch}%
{#1}\small\normalsize} \spacingset{1}

% -------------------------------------------------------------------------
% Title and authors

% Title for both blinded and unblinded versions...
\newcommand{\mytitle}{Optimal post-selection inference for sparse signals:
  a~nonparametric empirical Bayes approach}

\if0\blind
{
  \title{\bf \mytitle}
  
  \author{Spencer Woody\thanks{
       Corresponding author. Email to \texttt{spencer.woody@utexas.edu}}\hspace{.2cm}\\
    Department of Integrative Biology,\\
  The University of Texas at Austin \\
    and \\
    Oscar Hern\'an Madrid-Padilla \\
    Department of Statistics, The
  University of California, Los Angeles \\
  and \\
    James G. Scott \\
    Department  of Information, Risk, and Operations Management, \\The University of Texas at Austin}
  
%   More compact way
  
%   \author[1 2]{Carlos Carvalho}

% \author[1]{Jared Murray}

% \author[2]{Spencer Woody\thanks{Corresponding author. Email to
%     \texttt{spencer.woody@utexas.edu}}}

% \affil[1]{Department of Information, Risk, and Operations Management, University~of~Texas~at~Austin}

% \affil[2]{Department of Statistics and Data Sciences, University~of~Texas~at~Austin}

  \maketitle
} \fi

\if1\blind
{
  \bigskip
  \bigskip
  \bigskip
  \begin{center}
    {\LARGE\bf mytitle}

    % Optionally, if you want the date too

    % \bigskip
    % \today
    
\end{center}
  \medskip
} \fi

% -------------------------------------------------------------------------
% Abstract and keywords

\begin{abstract}
  Many recently developed Bayesian methods have focused on sparse signal detection.  However, much less work has been done addressing the natural follow-up question: how to make valid inferences for the magnitude of those signals after selection.  Ordinary Bayesian credible intervals suffer from selection bias, as do ordinary frequentist confidence intervals, owing to the fact that the target of inference is chosen adaptively.  Existing Bayesian approaches for correcting this bias produce credible intervals with poor frequentist properties, while existing frequentist approaches require sacrificing the benefits of shrinkage typical in Bayesian methods, resulting in confidence intervals that are needlessly wide.  We address this gap by proposing a nonparametric empirical Bayes approach for constructing optimal selection-adjusted confidence sets.  Our method produces confidence sets that are as short as possible on average, while both adjusting for selection and maintaining exact frequentist coverage uniformly over the parameter space.  Our main theoretical result establishes an important consistency property of our procedure: that under mild conditions, it asymptotically converges to the results of an oracle-Bayes analysis in which the prior distribution of signal sizes is known exactly.  Across a series of examples, the method outperforms existing frequentist techniques for post-selection inference, producing confidence sets that are notably shorter but with the same coverage guarantee.
  %\response{Use the \texttt{{\textbackslash}reponse\{\}} macro for new additions. Use} \sw{\texttt{{\textbackslash}sw}}  \jgs{\texttt{{\textbackslash}jgs\{\}}}  \response{and}  \omp{\texttt{{\textbackslash}omp\{\}}} \response{macros for your comments.}
\end{abstract}

\noindent%
\vfill

% -------------------------------------------------------------------------
% Set spacing for main text; don't change this

\newpage
\spacingset{1.45} % DON'T change the spacing!

% -------------------------------------------------------------------------
% Main body

%!TEX root = ../main.tex

\section{Introduction}
\label{sec:introduction}

This paper proposes a nonparametric empirical Bayes approach to post-selection inference with an exact coverage guarantee.  Our framework and theoretical results here are both very general, but we focus primarily on the commonly encountered setting of inference for a sparse vector of Gaussian means, where $y_i \sim N(\theta_i, \sigma^2)$, and where most $\theta_i$ are zero or negligibly small.  This model, although simple, is ubiquitous in modern statistical practice.

% For example, $y_i$ may
%represent the observed log-fold change in expression level for one
%gene between two conditions; it may quantify the statistical
%significance of a single voxel in an fMRI study; or it may measure
%differential methylation at a particular chromosomal location.

Much methodological work in this setting has focused on the question of how to find significantly nonzero $\theta_i$'s.  Many Bayesian approaches have been based on the two-groups model \citep[e.g.][]{efrontibshirani2001, ScottBerger2006, efron:2008}, where the prior is a mixture of a point mass at zero and some distribution of nonzero signals, or on global-local shrinkage priors \citep{Polson:Scott:2010a} such as the horseshoe \citep{horseshoe} or the Bayesian lasso \citep{bayeslasso}.  But regardless of the particular prior used, an important question that is not adequately addressed in this literature is how to quantify the magnitude of those signals selected for inference after the detection procedure is applied.  Previous scholars have differed on whether this poses a problem if one adopts a pure Bayesian perspective \citep[c.f.][]{Dawid1994,saBayes}.  However, \response{even if one does correctly adjust posterior inference for selection when necessary, using posterior credible interavls} certainly presents a very stark problem from a frequentist perspective.  It is widely known that Bayesian credible sets in general are not valid frequentist confidence sets, except under very specific ``matching'' priors \citep[e.g.][]{Ghosh}, \response{but rather only retain coverage on average with respect to the prior}.  Furthermore, in post-selection settings, the departure from nominal coverage is much more severe than is commonly appreciated.  For example, we present in \S\ref{sec:toy-example} a simple case where, under a plausible selection mechanism, the coverage of 90\% Bayesian credible sets is actually less than 50\% for very strong signals.

There is, of course, an active line of recent work on purely frequentist solutions to the post-selection inference problem \citep[e.g.][]{FCR,lee2016,Reid2015}.  
% For instance, \cite{FCR} provide confidence intervals to meet a false coverage rate,  \cite{lee2016} provide confidence intervals for coefficients in a lasso \citep{Lasso} , and \cite{Reid2015} introduce selection-adjusted confidence intervals which maintain the marginal nominal coverage rate.  
However, these adjustment procedures do not borrow information across components of the $\theta$ vector, and as a result, they produce confidence sets that are needlessly wide.  This is particularly problematic in post-selection inference because selective confidence intervals are necessarily wider than their ordinary (non-selective) counterparts.  \response{There have been several attempts to create shorter selection-adjusted frequentist intervals \citep{zhong2008bias,weinstein2013selection}, but none of these methods have explicitly incorporated Bayesian ideas to improve the efficiency of inference.}

Thus there is a major unmet need for inferential procedures that: (i) correctly adjust for the effects of selection; (ii) maintain valid frequentist coverage, uniformly across the whole parameter space; and (iii) produce confidence sets that are as short as possible.  The approach we propose here has all three of these desirable features.  As we will illustrate, while our procedure is frequentist in nature, its efficiency gains arise from Bayesian thinking: that is, from positing the existence of a prior that describes the sparsity pattern in the data set, and then estimating that prior nonparametrically using modern empirical Bayes tools.

% We
% refer to the method as ``saFAB,'' for selection-adjusted FAB
% inference.

% \begin{align}
% \label{eqn:joint-joint}
%   p_S^{(J)}(\theta, y) = \frac{\pi(\theta) \cdot f(y \mid \theta) } {\pr(S  )} \cdot \mathbf{1}(y \in S),
% \end{align}

% The selection-adjusted posterior for $\theta$, given any $y$ that
% falls in $S$, is
% \begin{align}
% \pi^{(J)}_S(\theta \mid y) = \frac{\pi(\theta) \cdot f(y \mid \theta)}
%   {\pr(S  )}  \, ,\label{eq:joint-posterior}  
% \end{align}
% and the marginal is
% \begin{align}
% \label{eq:joint-marginal}
%   \begin{split}
% m^{(J)}_S(y)  &= \int \mathbf{1}(y \in S) \pi(\theta) \ f(y  \mid 
%              \theta) / \Pr(S  ) \ \mathrm{d}\theta \\
%              &= m(y) \cdot \mathbf{1}(y \in S) / \Pr(S  ) \, ,
%   \end{split}
% \end{align}

% For conditional selection,
% \begin{align}
%   \label{eq:conditional-joint}
% p_S^{(C)}(\theta, y)  = \frac{\pi(\theta) \cdot f(y \mid \theta) } {\pr(S   \mid \theta)} \cdot \mathbf{1}(y \in S) \, .
% \end{align}

% The posterior is
% \begin{align}
%   \label{eq:conditional-posterior}
%   \pi^{(C)}_S(\theta \mid y) \propto \frac{\pi(\theta) \cdot f(y \mid
%   \theta)} {\pr(S \mid \theta)} 
% \end{align}
% and the marginal is
% \begin{align}
% \label{eq:conditional-marginal}
%     m_S^{(C)}(y) = \mathbf{1}(y \in S) \int \pi(\theta) \ f(y  \mid  \theta) /
%     \Pr(S  \mid  \theta) \ \mathrm{d}\theta\, ,
% \end{align}

% Frequentist methods involve using the selection-adjusted likelihood,

% TEXShop 
% !TEX root = ../main.tex

% Emacs
%%% Local Variables:
%%% mode: latex
%%% TeX-master: "../main"
%%% End:

%!TEX root = ../main.tex

\section{Efficient selection-adjusted inference}
\label{sec:safab}

\subsection{Exchangeable means with known prior}    

In this section we describe our procedure under the simplifying assumption that the signals $\theta_i$ arise from a known prior distribution $\pi(\theta)$.  In \S\ref{sec:nonp-proc}, we will come to the core of our proposal, when we describe a method for the more realistic setting of unknown $\pi(\theta)$.

Our work builds off the ``frequentist assisted by Bayes'' (FAB) framework, which was articulated in its earliest form by \cite{pratt1963} and substantially extended by \cite{YuHoff}.  This technique was originally proposed for constructing confidence intervals for group-level means in hierarchical normal models, and has also been extended to constructing confidence intervals for coefficients in a linear regression by \cite{HoffYu}.  Here we generalize the FAB procedure to produce Bayes-optimal confidence sets under post-selection inference.  We call this procedure selection-adjusted FAB (saFAB).

Suppose we observe $y_1, \ldots, y_n$ \response{independently} from the parametric model $y_i \sim f(y_i; \theta_i)$, and we consider inference \response{only for units $i$ such that} $y_i \in S$ for some region $S \subset \mathbb{R}$, which we refer to as the selection region or selection event.  Conditional on this selection event, $y$ has the truncated density function
\begin{align}
  \label{eq:sa-likelihood} f_S(y ; \theta) &= f(y ; \theta) \cdot
\mathbf{1}(y \in S)/ \int_S f(y ; \theta) \mathrm{d} y,
\end{align}
with cumulative density function $F_S(y ; \theta)$ and generalized quantile function $F_S^{-1}(p; \theta) = \inf_y \{y: F_S(y ; \theta) \geq p\}$.  Using \eqref{eq:sa-likelihood}, one may construct confidence sets that condition on the selection event $y \in S$ via the well-known procedure of inverting a family of $\alpha$-level tests of point null hypotheses of the form $H_0: \theta = \theta_0$ for all $\theta_0 \in \mathbb{R}$.  The key decision that arises in this procedure, and that we will discuss in detail below, is how to construct this family of tests.

We first remark that the standard choice of an equal-tailed test is the universally most powerful unbiased (UMPU) test and forms the basis of the method proposed by \cite{Fithian}.  In particular, suppose that we condition on $y \in S$, so that $y \sim f_S(y ; \theta)$.  Then the UMPU test of $H_0: \theta = \theta_0$ has acceptance region
\begin{align}% \label{acceptanceUMPU}
  A^{S}(\theta_0) = \{y: F_S^{-1}(
\alpha/2 ; \theta_0) \leq y \leq F_S^{-1}(1 - \alpha/2 ; \theta_0)
\}. \nonumber
\end{align}
As shown by \cite{Fithian}, inverting this family of acceptance regions gives the ($1-\alpha$)-level uniformally most accurate unbiased (UMAU) selection-adjusted confidence interval, of the form $C^{S}(y) = \{ \theta: y \in A^S(\theta) \}$. It is clear this interval will cover the true $\theta$ with the correct probability, conditional on the selection event $y \in S$, because $\pr_{\theta}(\theta \in C^S(y) \mid y \in S) = \pr_{\theta}(y \in A^S(\theta)) = 1 - \alpha$.

However, following the logic that both \cite{pratt1963} and \cite{YuHoff} apply to the task of ordinary (non-selection-adjusted) inference, we need not restrict ourselves to confidence sets based on the unbiased test using equal-tail probability regions in $F_S(y ; \theta)$. In fact, we may generally choose a biased test with acceptance region. 
\begin{align}
  \label{acceptance} A^S_w(\theta_0) &= \{y:
F^{-1}_S(\alpha w ; \theta_0) \leq y \leq F^{-1}_S(\alpha w + 1 -
\alpha ; \theta_0) \} \, ,
\end{align}
where $w \in [0,1]$ controls how $\alpha$, the total probability mass that falls outside the acceptance region, is split between the two tails of $f_S(y ; \theta)$. This leads to confidence sets of the form $C^S_w(y) = \{ \theta: y \in A^S_w(\theta) \}$, which will retain nominal coverage by construction. We note that the choice $w=0.5$ recovers the equal-tailed UMPU test and thus the confidence sets from \cite{Fithian}; any other choice will put different probabilities in the left versus right tail.  In fact, we are free to choose a different value of $w$ for each $\theta_0$ in Eq.~\eqref{acceptance}, and may tune these choices to reflect our prior knowledge.  Specifically, we will find a form of $w$ to give the shortest confidence sets for $\theta$, in expectation under a prior $\pi(\theta )$.  The result will be a function $w: \mathbb{R} \rightarrow [0, 1]$ that gives a different optimal $w$ for each $\theta$, yielding a family of acceptance regions $A^S_{w(\theta)}$ and confidence set procedure $C^S_{w(\theta)}$.  Following \cite{YuHoff}, we call $w(\theta)$ the spending function, as it dictates how the acceptance region allocates, or spends, the Type I error rate~$\alpha$.

The Bayes-optimal spending function is found as follows.  First, \response{following \cite{pratt1961length} and \cite{YuHoff}}, define the frequentist risk $ R(\theta; w)$ of a confidence-set procedure for $\theta$ to be the expected Lebesgue measure of that confidence region under $f_S(y; \theta)$.  That is,
\begin{align*} R(\theta; w) & = \int \int \mathbf{1}(y \in
A^S_w(\tilde{\theta})) f_S(y ; \theta) \mathrm{d} \tilde{\theta}
\mathrm{d}y.
\end{align*}
Then, upon introducing a prior $\theta \sim \pi(\theta)$, we may compute the Bayes loss of a confidence set $C^S_w$ procedure as
\begin{align} L(\pi, w) &=\int R(\theta; w) \pi(\theta)
\mathrm{d}\theta
    = \int \left[ \int \int \mathbf{1}(y \in A^S_w(\tilde{\theta}))
f_S(y ; \theta) \mathrm{d}\tilde{\theta} \mathrm{d}y \right]
\pi(\theta) \mathrm{d}\theta \nonumber \\
    & = \int \left[ \int \int \mathbf{1}(y \in A^S_w(\tilde{\theta}))
f_S(y ; \theta) \pi(\theta) \mathrm{d}y\mathrm{d}\theta \right]
\mathrm{d} \tilde{\theta}
               = \int \pr(Y \in A^S_w(\tilde{\theta}))
\mathrm{d}\tilde{\theta}. \label{eq:bayesrisk}
\end{align}
The final integrand in Eq.~\eqref{eq:bayesrisk} is the marginal probability that $y$ falls in the acceptance region, integrating out $\theta$ under the prior.  Intuitively, confidence procedures that yield shorter intervals, on average under the marginal distribution of selected observations, have smaller Bayes loss.  Let $m_S(y)$ denote the density of this marginal distribution, whose form we discuss in the following subsection, with corresponding cumulative density function $M_S(y)$.  We may rewrite the integrand in \eqref{eq:bayesrisk} as
\begin{align}\label{eq:objectivefun}
  H(w; \theta) \equiv \pr(Y \in
A^S_w(\theta)) = M_S\left[ F_S^{-1}(\alpha w + 1 - \alpha ; \theta)
\right] - M_S\left[ F_S^{-1}(\alpha w ; \theta) \right].
\end{align}
This defines an objective whose minimization gives the Bayes-optimal spending function,
\begin{align} w^\star(\theta) = \arg \min_{w \in [0, 1]} H(w;
\theta). \nonumber
\end{align}
Generally, there is no closed-form solution to solve this optimization problem.  Therefore we implement a numerical approach whereby the spending function is approximated pointwise by solving the minimization problem along a grid of $\theta$ values.  After the optimal spending function is calculated, the Bayes-optimal confidence set for selected observations is obtained by inverting the resulting acceptance region, $C^S_{w^\star(\theta)}=\{\theta: y \in A^S_{w^\star(\theta)}\}$.

\response{We note that $w(\theta)$ must be a monotonic function to ensure that the produced confidence sets are intervals.  This is sometimes not the case, as in the toy example presented later in this section.  Thus, the returned confidence set will occasionally be a set of disjoint intervals, in which case the size of the set is equal to the Lebesque measure.  However, this rarely happens in all our applications.  If such a confidence set is produced, one may create an interval by taking the minimum and maximum of the set, and then the size of this interval would be given by the difference.  Such a procedure will produce confidence intervals with conservative coverage and which will often still be more efficient than the UMAU intervals. }

\subsection{Influence of the selection mechanism on the spending
function}
\label{sec:determ-spend-funct}

We observe from Eq.\eqref{eq:objectivefun} that calculating the optimal spending function requires knowing two parts of the overall model: the truncated sampling model $f_S(y ; \theta)$ and the marginal density of selected signals $m_S(y)$.  As we shall see, this marginal density depends both on the prior $\pi(\theta)$ as well as the specific mechanism by which selection occurs.

A very useful distinction between two different types of selection mechanisms---joint selection versus conditional selection---was drawn by \cite{saBayes}.  This distinction is best explained by example.  To understand joint selection, imagine a genomics study that seeks to understand the differences in gene expression across two experimental conditions.  Suppose the data arise as follows.  For each gene $i = 1, \ldots, N$: (i) draw $\theta_i \sim \pi(\theta)$, representing a true effect size for gene $i$; (ii) observe data $y_i \sim N(\theta_i, 1)$, representing the observed effect size for gene $i$, (iii) select the $(y_i, \theta_i)$ pairs where $y_i \in S$ (e.g.~$|y_i| > 2$), and perform inference only for those selected $\theta_i$'s.  Under such a scenario, \response{the selected} $y$ and $\theta$ have the joint distribution
\begin{align}
\label{eqn:joint-joint} p_S^{(J)}(\theta, y) = {\pi(\theta) \cdot f(y
; \theta) } \cdot \mathbf{1}(y \in S) /{\pr(S )},
\end{align}
where $\pr(S) = \int \int_S f(y; \theta) \pi(\theta) \dee y \dee \theta$ is the marginal probability of selection.  In this case, the marginal density of selected signals is proportional to the ordinary marginal density $m(y)=\int f(y;\theta)\pi(\theta) \dee \theta$, truncated to $S$:
\begin{align}
\label{eq:joint-marginal}
  \begin{split} m^{(J)}_S(y) &= \int \mathbf{1}(y \in S) \pi(\theta) \
f(y; \theta) / \pr(S ) \ \mathrm{d}\theta = m(y) \cdot \mathbf{1}(y
\in S) / \pr(S ).
  \end{split}
\end{align}
We use the term ``joint selection'' rather than Yekutieli's term ``random parameter selection'' to describe such a setting, to emphasize that the selection mechanism applies to $(y_i, \theta_i)$ pairs jointly.

Now turning to conditional selection, imagine a scientific field with many open questions ($i = 1, 2, \ldots$), where journals publish results only for observed effect sizes $y \in S$.  For each open question $i$, published results are generated as follows: (i) sample $\theta_i \sim \pi(\theta)$, representing the true effect size for question $i$; (ii) many labs ($k = 1, 2, \ldots$) observe $y_i^{(k)} \sim N(\theta_i, \sigma^2)$; (iii) the first lab that observes $y_{i}^{(k)} \in S$ publishes its results.  Now let $y_i$ denote the published estimate for $\theta_i$; by construction, this is a sample from $N(\theta_i, \sigma^2)$ truncated to $S$, taken by rejection sampling.  We refer to this mechanism as ``conditional selection'', rather than Yekutieli's term of ``fixed parameter'' selection, to emphasize that the selection mechanism applies to draws for $y_i$, conditional upon a specific value of $\theta_i$, which is itself assumed to be a random draw from a prior.

Under conditional selection, the joint distribution of $(\theta, y)$ for selection signals is now
\begin{align}
  \label{eq:conditional-joint} p_S^{(C)}(\theta, y) = {\pi(\theta)
\cdot f(y; \theta) } \cdot \mathbf{1}(y \in S) /{\pr(S \mid \theta)}
\, ,
\end{align}
where $\pr(S \mid \theta) = \int_S f(y; \theta) \dee \theta$ is the probability of selection conditional on $\theta$.  This implies that the marginal density of the data is
\begin{align}
\label{eq:conditional-marginal} m_S^{(C)}(y) = \mathbf{1}(y \in S)
\int \pi(\theta) \ f(y \mid \theta) / \pr(S \mid \theta) \
\mathrm{d}\theta.
\end{align}
Unlike in the case of joint selection, this marginal is no longer equivalent to the usual marginal density truncated to $S$, because of the term $\pr(S \mid \theta)$ appearing inside the integral over $\theta$.

Thus we see that the optimal spending function depends on four factors: the parametric model $f(y; \theta)$, the selection set $S$, the prior $\pi(\theta)$, and the selection mechanism itself.

\subsection{Toy example}
\label{sec:toy-example}

\begin{figure}[t!]
  \centering
\includegraphics[width=0.65\textwidth]{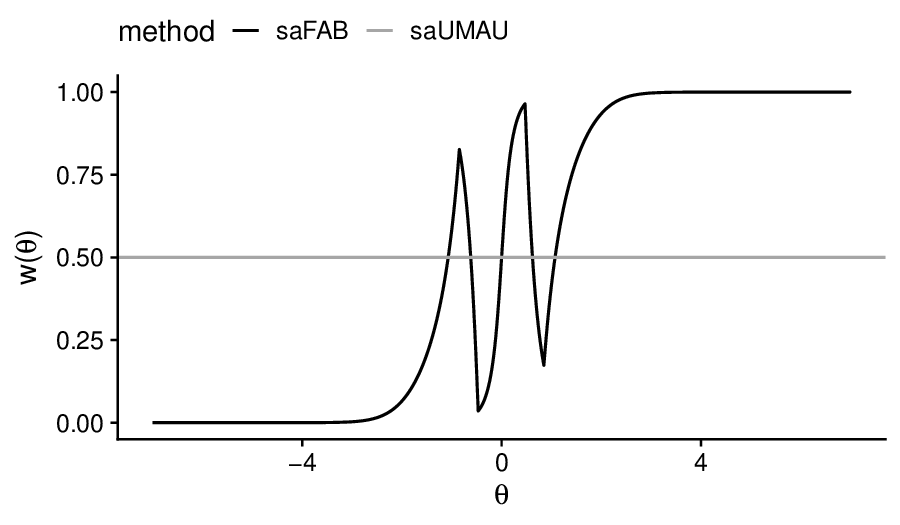}
\caption{\label{fig:toy-spend-fun}
  Spending function from the toy example, which determines the degree of bias in acceptance region calculations as a function of $\theta$.  Inversion of these acceptance regions gives the Bayes-optimal confidence sets.  The spending function depends on (i) the sampling distribution of observations, (ii) the selection set, (iii) the selection mechanism, and (iv) the prior for signals.
}
\end{figure}

To demonstrate the mechanics and advantages of our method, we first show how it performs on a toy example where $\theta$ follows the two-groups model,
\begin{align}\label{eq:two-groups-normal}
  \pi(\theta) &= p \cdot {N}(\theta; 0, \tau^2) + (1 - p) \cdot \delta_0(\theta),
\end{align}
a mixture of a point mass at 0 and a zero-centered Gaussian with variance $\tau^2$.  For this example we set $p=0.1$ and $\tau^2 = 3$. We generate $n = 10,000$ samples of $\theta_i$ from this distribution, and then sample each $y_i$ from ${N}(\theta_i, 1)$. The selection region is chosen to be $S = \{y: |y| > 2 \}$ and we operate under the joint selection mechanism.  That is, $(\theta_i, y_i)$ are sampled jointly and a confidence set is constructed for $\theta_i$ only if $|y_i| > 2$.

Figure~\ref{fig:toy-spend-fun} shows the resulting optimal spending function for this particular combination of prior, sampling model, selection region, and selection mechanism.  This spending function defines the family of selective hypothesis tests to create the Bayes-optimal confidence sets.  For comparison, we also calculated the UMAU confidence sets from \cite{Fithian} which correspond to the flat spending function $w_\text{saUMAU}(\theta) = 1/2$.

\begin{figure}[t!]  \centering
\includegraphics[width=0.7\textwidth]{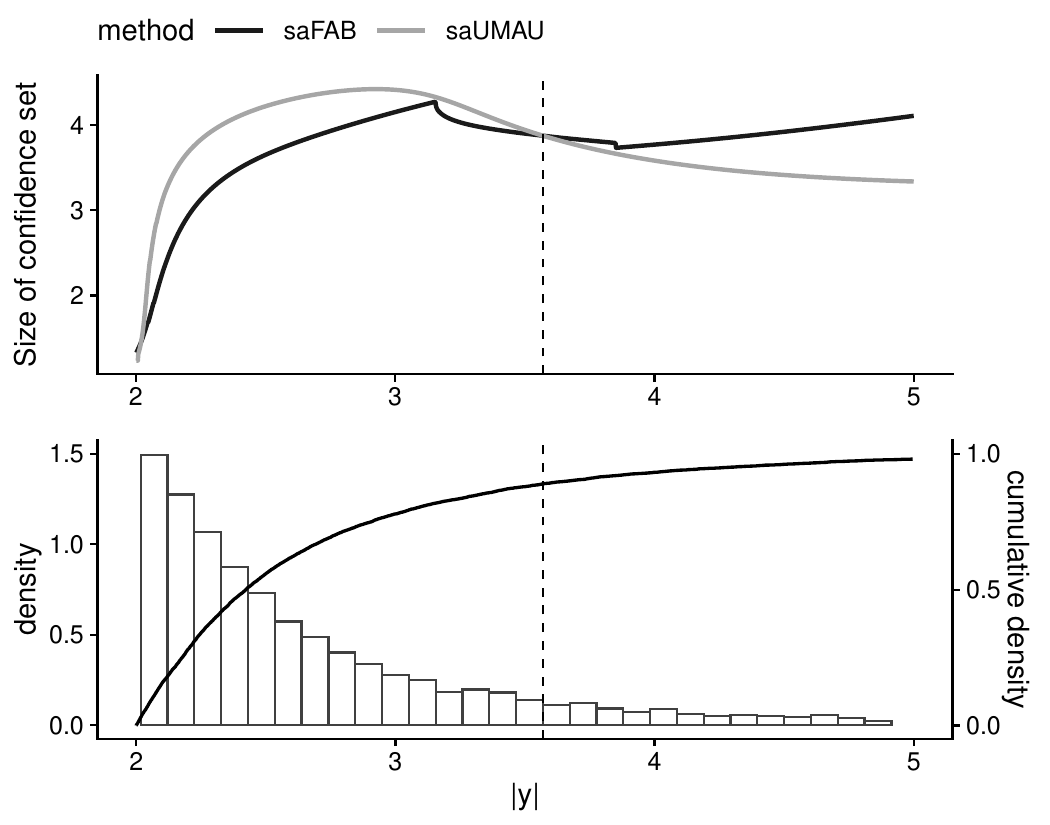}
\caption{\label{fig:toy-example-length}
  Comparison of sizes of selection-adjusted UMAU confidence sets (saUMAU) and our selection-adjusted FAB confidence sets as a function of observation size.  For most observations (around 90\% of the marginal density), saFAB returns a smaller confidence set, resulting in confidence sets that are 12\% smaller on average.
}
\end{figure}

Figure~\ref{fig:toy-example-length} shows the sizes of each procedure's confidence sets $C(y)$ as a function of $y$.  For reference, we have also plotted the marginal distribution of $y$ for the selected signals.  Notably, the saFAB procedure gives a smaller confidence set for values of $y$ that are most frequently observed: for approximately 90\% of the observations in this example, saFAB returns a smaller confidence set.  For values of $y$ larger than approximately 3.6, saFAB give a wider confidence set than the intervals from \cite{Fithian}, but observing a $y_i$ this large is relatively rare.  As a result, the saFAB intervals are 12\% shorter on average than the UMAU intervals.

\begin{figure}[t!]  \centering
\includegraphics[width=0.9\textwidth]{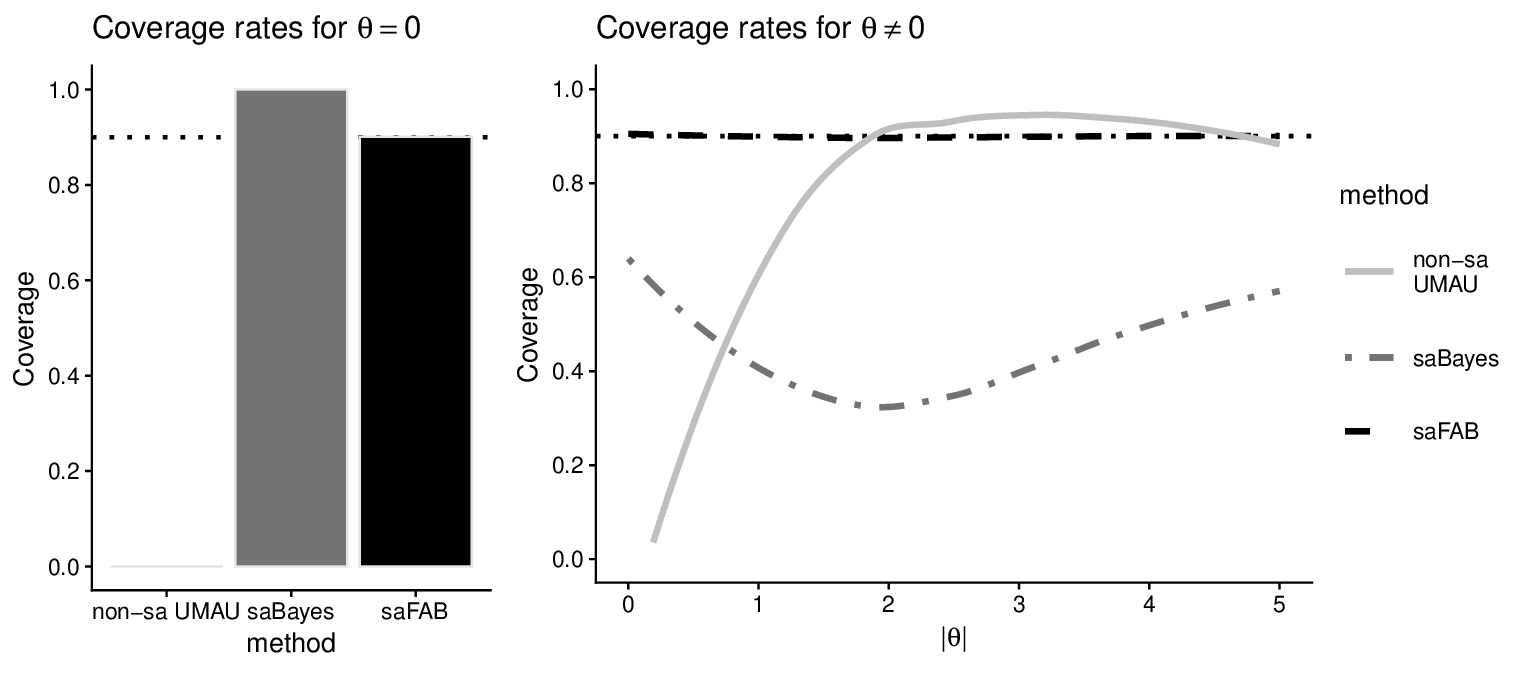}
\caption{\label{fig:toy-example-coverage}
  Results for toy example, comparing coverage rates for non-selection adjusted confidence sets (non-sa UMAU), selection-adjusted Bayesian posterior credible intervals (saBayes), and our selection-adjusted FAB confidence sets (saFAB).  We delineate the two cases of $\theta = 0$ (left) and $\theta \neq 0$ (right).  Non-selection adjusted confidence sets have poor coverage throughout and selection-adjusted credible intervals only have correct coverage on average across the prior.  saFAB confidence sets have uniform nominal coverage.
}
\end{figure}

Finally, Figure~\ref{fig:toy-example-coverage} shows the coverage of our saFAB method in comparison to two alternative methods: the non-selection adjusted confidence sets (``non-sa UMAU'') and selection-adjusted Bayesian credible intervals (``saBayes''). \response{Following \cite{saBayes}, the selection-adjusted posterior distribution is identical to the ordinary posterior distribution because of the mechanism of joint selection.} Since the $\theta_i$ are drawn from the mixture in Eq.~\eqref{eq:two-groups-normal}, we consider the two cases of $\theta = 0$ and $\theta \neq 0$.  The unadjusted confidence intervals have poor coverage, with nearly zero coverage for $\theta$ values equal to and near 0, owing to selection bias.  On the other hand, the selection-adjusted Bayesian credible intervals have correct coverage on average across all signals, but this coverage is highly non-uniform in $\theta$.  In particular, since the credible interval nearly always includes 0 due to the presence of the point mass in the posterior, the coverage for $\theta=0$ is nearly 1, while coverage is much lower for all other values of $\theta$.  \response{In the supplemental material, we show how this problem of poor coverage of credible intervals for nonzero $\theta$ is exacerbated when even greater prior mass is assigned for $\theta = 0$.  }  Our saFAB procedure, on the other hand, provides nominal coverage uniformly across the entire parameter space.  Although not shown, the usual selection-adjusted UMAU confidence sets of \cite{Fithian} also exhibit the same uniform coverage; they are simply wider on average.

% TEXShop % !TEX root = ../main.tex

%%% Local Variables:
%%% mode: latex
%%% TeX-master: "../main"
%%% End:

%!TEX root = ../main.tex

\section{Nonparametric empirical Bayes procedure}
\label{sec:nonp-proc}

\subsection{Estimating the prior}
\label{sec:basic-approach-main}

In the previous section we introduced the saFAB procedure for situations where the prior $\pi(\theta)$ had a known parametric form.  In cases where $\pi(\theta)$ is not known, however, it may not be desirable to specify a parametric form or elicit a subjective choice. For this situation, we propose a method for post-selection inference where a prior need not be fully specified but rather is estimated from the data using a nonparametric empirical Bayes approach.  Throughout, we still assume that both the sampling model $f(y; \theta)$ and the selection mechanism are known, and that the selection rule $S$ has also been pre-specified.  Our main theoretical result establishes that, under mild regularity conditions and with selection occurring jointly on $(\theta, y)$, our nonparametric method gives a consistent estimate of the true optimal spending function.  In addition, our empirical results show that this data-driven strategy for constructing a spending function offers substantial efficiency improvements over the UMAU intervals, while losing very little compared to a hypothetical oracle-Bayes analysis in which the prior is known.

% "Subdensity"

Recall that the optimal spending function $w^\star(\theta)$ is calculated by minimizing the objective function $H(w; \theta)$ in Eq.~\eqref{eq:objectivefun}, which involves the marginal density of selected observations $m_S(y)$.  This marginal density in turn depends on both the selection mechanism, which is known, as well as the prior, which is unknown.  The intuition of our approach is that if we are able to specify a prior whose corresponding marginal distribution matches the data well, we should also be able to recover a good estimate of the optimal spending function.  Our nonparametric empirical Bayes approach exploits this fact by using a plug-in estimate $\hat{\pi}(\theta)$ for the prior distribution.

One challenge that arises in constructing such an estimate $\hat{\pi}(\theta)$ is that it must adapt to the data, yet still ensure valid post-selection sets.  As implied by Proposition 7 from \cite{YuHoff}, in order to retain nominal coverage of the procedure, the spending function---and therefore the estimate of the prior---cannot depend on the same data that we use to construct confidence sets.  Otherwise, there is the potential for selection bias to be re-introduced via the mechanism of an overly optimistic prior.  \response{As an example, consider the data-dependent prior that concentrates all its probability mass on the observed value of $y$: $\pi(\theta_i) = \delta_{y_i}$.  It is easy to show that, if we use our procedure to construct the optimal spending function under such an aggressively data-dependent prior, the resulting confidence sets will not retain nominal coverage.}

To avoid a more subtle version of this pitfall, in practice we use a $K$-fold data-splitting approach, whereby the data are split into $K$ non-overlapping subsets $y_1,\ldots,y_K$.  To construct confidence sets for the data points in fold 1, we first use the data in folds 2 through $K$ to form an estimate of the prior $\hat{\pi}(\theta)$.  This gives us an estimate of the optimal spending function $\hat{w}(\theta)$, which we use to construct Bayes-optimal saFAB confidence sets for the data points in fold 1.  \response{Note that the data in the other folds are used only to estimate the spending function, but the actual confidence interval for a single $\theta_i$ is constructed using this spending function and only the single observation $y_i$.}  We then repeat this process for each fold, always holding out the data from that fold when we estimate the prior.  This ensures that the independence condition from Proposition 7 of \cite{YuHoff} is met.  To actually form the estimate $\hat{\pi}(\theta)$, we use the method of predictive recursion from \cite{Newton02ona}.  We provide details of this approach in Algorithm~\ref{al:pr} of Appendix~\ref{sec:predictive-recursion}.

\subsection{Constructing the spending function}

We now assume that $\hat{\pi}(\theta)$ is given, and we use it to obtain an estimate of the optimal spending function $\hat w^\star(\theta)$ as follows.  First, from the estimated prior $\hat{\pi}(\theta)$, we derive the corresponding marginal distribution of the data under selection, $\hat m_S(y)$, which is calculated by substituting $\hat \pi(\theta)$ in for $\pi(\theta)$ in either Eq.~\eqref{eq:joint-marginal} or Eq.~\eqref{eq:conditional-marginal}, depending on the selection mechanism.  That is, the estimated selection-adjusted marginal is either
\begin{align}
  \label{eq:joint-marginal-est}
 \hat m^{(J)}_S(y)  &= \int \mathbf{1}(y \in S) \hat \pi(\theta) \ f(y; 
             \theta) / \hat \pr(S  ) \ \mathrm{d}\theta = \hat m(y) \cdot
             \mathbf{1}(y \in S) /  \hat \pr(S  ). 
\end{align}
for joint selection, where $\hat\pr(S)=\int \int_S f(y;\theta) \hat \pi(\theta) \dee y \dee \theta$ and $\hat m(y) = \int f(y; \theta) \hat \pi(\theta)\dee \theta$, or
\begin{align}
  \label{eq:conditional-marginal-est}
      \hat m_S^{(C)}(y) = \mathbf{1}(y \in S) \int \hat\pi(\theta) \ f(y  \mid  \theta) /
    \pr(S  \mid  \theta) \ \mathrm{d}\theta\, ,
\end{align}
for conditional selection.  This then defines the surrogate objective function
\begin{align}\label{eq:objectiveapprox}
  \hat H(w; \theta) 
  % = \pr(y \in A^S_w(\theta))
  = \hat M_S\left[ F_S^{-1}(\alpha w + 1 - \alpha ; \theta) \right] -
  \hat M_S\left[ F_S^{-1}(\alpha w ; \theta) \right],
\end{align}
which serves as a proxy for the true (unknown) objective function $H(\theta; w)$ in Eq.~\eqref{eq:objectivefun}.  Minimization of this objective function in turn gives the estimated optimal spending function $\hat w^\star(\theta) = \arg \min_{w \in [0, 1]} \hat H(w; \theta)$.  We then use this estimate of the optimal spending function to calculate confidence sets, in the manner described in \S\ref{sec:safab}.Œ

% \jgs{All of these things need to be inside the theorem.  Golden rule
%   of stating a theorem is that you must state conditions necessary for
%   the theorem to be true \textit{inside} the theorem itself.}

%To recap, the full empirical-Bayes procedure is as follows
%\begin{enumerate}
%\item Specify the sampling model as $f(y ; \theta)$, and the selection
%  region $S$, which together determine the selection-adjusted sampling
%  model $f_S(y ; \theta)$.  Posit the selection mechanism as either
%  occurring jointly on $(\theta, y)$, or conditionally on $\theta$.
%\item Split the data into $K$ folds $y_1,\ldots,y_K$.
%\item For fold 1, use the 
%\item With the data $y_{k'}$, estimate the prior
%  $\hat \pi(\theta)$ using predictive recursion (see
%  Algorithm~\ref{al:pr} in Appendix~\ref{sec:predictive-recursion}).
%\item Calculate the estimated selection-adjusted marginal
%  $\hat m_S(y)$ using the selection-adjusted sampling model
%  $f_S(y; \theta)$ and the estimated prior $\hat{\pi}(\theta)$.
%\item Using $f_S(y ; \theta)$ and $\hat m_S(y)$, construct the estimated
%  optimal spending function $\hat{w}(\theta)$ by minimizing the
%  objective function $\hat H(w; \theta)$.
%\item Use $\hat{w}^\star(\theta)$ to construct the family of biased tests
%  for each value of $\theta$.
%\item Invert this family of biased tests to get the optimal selective
%  confidence sets for each $y \in y_{k}$.
%\item Repeat Steps 2 through 7 for each $k \in \{1, ..., K\}$. 
%\end{enumerate}

\subsection{Main consistency result}

The key statistical question that arises from our procedure is whether the minimizer of the surrogate objective $ \hat H(w; \theta) $ is a good estimate for the minimizer of the true, unknown objective $ H(w; \theta) $.  We now turn to our main theoretical result, which establishes that, under mild conditions, the answer is yes, at least in the case of joint selection.  Our proof builds upon the foundation laid by \cite{Tokdaretal}, who give conditions under which predictive recursion will yield an estimate $\hat m(y)$ that converges to the true marginal density $m(y)$.  From their result, it is easy to demonstrate that the estimated selection-adjusted marginal density for joint selection, $\hat m_S^{(J)}(y)$ in Eq.~\eqref{eq:joint-marginal-est} also converges to the true density $m^{(J)}_S$ in Eq.~\eqref{eq:joint-marginal}.  Our theorem hinges on using this fact to show that the worst-case distance from any minimizer $\hat w^\star(\theta)$ of the surrogate objective function $\hat H(\cdot;\theta)$ to the set of minimizers of $H(;\theta)$ converges to zero in probability; see Eq.~\eqref{eqn:con1} below. \response{Thus, for  any minimizer $\hat{w}^\star(\theta)$ of  $\hat  H(\cdot;\theta)$ there exists a minimizer $w^\star(\theta)$ of $H(\cdot;\theta)$ which is close to   $\hat{w}^\star(\theta)$. 
	Furthermore, we show that the value of $H(\cdot;\theta)$ at any minimizer of $\hat H(\cdot;\theta)$ converges to the optimal value of $H(\cdot;\theta)$. These results require the  same conditions in \cite{Tokdaretal} that are necessary to ensure the Kullback-Leibler convergence of the predictive-recursion estimator.   The explicit assumptions are given in Appendix \ref{sec:asumptions}. The first  condition has to do with the choice of the weights in the predictive recursion algorithm, whereas the remaining  are regularity assumptions on the model. These  include an identifiability  condition of the mixing density when the likelihood is  $f(\cdot;\theta)$;  the requirement that $f(y;\theta)$ is bounded and continuous as a function of $\theta$;  and two  boundedness properties involving the behavior of  $f$.	As shown in  \cite{Tokdaretal}  these conditions all hold for the Gaussian model.}

\begin{theorem}
	%   \label{mytheorem}
	% \subsection{Theorem}
	\label{mytheorem} 
	%\response{ss}% except at two points $-t$ and $t$, for $t>0$
	\response{Suppose that the selection region is $S=\{y: |y| > t\}$ for some $t > 0$ with selection occurring jointly on $(\theta, y)$, and that $f(\cdot; \theta)$ has as support the real line.  Also suppose that the estimated prior $\hat \pi(\theta)$ is calculated using predictive recursion as outlined in Algorithm~\ref{al:pr} using a subset of the data that grows with $n$.  If Assumptions A1-A5 from   \cite{Tokdaretal} are met by  $f$  (see   Appendix \ref{sec:asumptions}), then for  a fixed  $\theta$  the following statements hold:}
	\begin{itemize}
		\item %Let  $\hat{\Omega}$ the set of minimizers of $\hat H(\cdot;\theta)$, and let 
		Let $\Omega_0$ be the set of minimizers of $H(\cdot;\theta)$, and $\hat{\Omega}$ the set of minimizers of $\hat H(\cdot;\theta)$. 
		%Thus,
		%% 
		%\begin{align*}
		%\Omega_0 =   \left\{
		%w:
		% \min_{\tilde w \in [0, 1]} H(\tilde{w};\theta) =      H(w;\theta)
		% \right\},
		%\end{align*}
		%and
		%\begin{align*}
		%\hat{	\Omega} =
		%\left\{
		% w:
		% \min_{\tilde w \in [0, 1]} \hat H(\tilde{w};\theta) = \hat H(w;\theta)
		%\right\}.
		%\%end{align*}
		\response{Assume that $\Omega_0 \neq \emptyset$. If   $q:=F_{S}(t,\theta)  \leq  \alpha$ and  $q/\alpha \notin  \Omega_0$ then we require that $\lim_{w \to (q/\alpha)^{+} } H(w;\theta) >  \min_{\tilde w \in [0, 1]} H(\tilde{w};\theta). $
			%\begin{equation}
			%	\label{eqn:extra}
			%	\end{equation}
			Then 
			% and  for  every $q \in \Lambda \backslash \Omega_0$, with $\Lambda$ the set of discontinuities of  $H(\cdot;\theta)$, we have that $\lim_{w \to q^{+}  } H(w;\theta) >  \min_{\tilde w \in [0, 1]} H(\tilde{w};\theta) $. Then  
			for every $\epsilon>0$ we have}
		\begin{align}
		\label{eqn:con1}
		\underset{n \to  \infty }{\lim}     \pr\left(
		\sup_{ \hat w^\star \in \hat{\Omega}  }
		\inf_{ w^\star \in \Omega_0  }
		\left\vert \hat w^\star - w^\star \right\vert \geq
		\epsilon
		\right)\, =\, 0.
		\end{align}
		\response{Thus,  the worst-case distance  from  $\hat{\Omega}$ to  $\Omega_0$    converges to zero in probability. }
		\item Almost surely, if $ \hat w^\star \in \hat{\Omega}$ then
		\begin{align}
		\label{eqn:con2}
		\underset{n \to \infty}{\lim}\,   H(\hat w^\star;\theta)  \,= \, H(w^\star;\theta),
		\end{align}
		for some  $w^\star \in \Omega_0$.
		% there exists  a random  subsequence $\{ w_{n_k}^\star \}  $  of  $\{  w_n^\star \}$  such that
		% \begin{align}
		% \label{eqn:con2}
		% w_{n_k}^\star   \,\rightarrow \,  w^\star,
		% \end{align}
		% 
		% for some $w^\star   \in  \Omega_0$. In fact,  all the limit points of  $\{w_n^\star\}$  belong  to $\Omega_0$.
	\end{itemize}
\end{theorem}

\begin{proof}
	See Appendix~\ref{sec:proof-main-result}.
\end{proof}

\response{  We notice that  when $ q	=F_S(t;\theta) \leq  \alpha$ and $q/\alpha \notin  \Omega_0$ the conclusion (\ref{eqn:con1})    requires an extra condition.  If such condition is violated, thus,
	% a  condition on the set of discontinuities of the function $\Lambda$. Importantly,  it is the case that $\vert H \vert  \in \{0,1,2\}$. If  for  some  $q\in \Lambda \backslash \Omega_0$ it holds that
	\begin{equation}
	\label{eqn:violation}
	\lim_{w \to (q/\alpha)^{+}  } H(w;\theta)  =  \min_{w\in [0,1]}  H(w;\theta), 
	\end{equation}
	then (\ref{eqn:con1})  will hold replacing  $\Omega_0$ with $\Omega_0 \cup \{q/\alpha\}$. This is reasonable since  when  (\ref{eqn:violation})  holds even minimization  of  $H$ over a finite  grid of points in $[0,1]$  could lead  to  points that are arbitrarily close to  $q/\alpha$ as the grid size grows. 
}

\response{An important  implication  of Theorem \ref{mytheorem} has to do with constructed confidence sets. In particular,  let  $\theta_1,\ldots,\theta_N$  fixed points  in the real line. If all the assumptions in Theorem \ref{mytheorem} hold for  each  $\theta_j$   and  $F_S(t;\theta_j)/\alpha$ is not a minimizer of $H(\cdot;\theta_j)$ for  all  $j =1,\ldots, N$, then for  each $j$ there exists a  $w^{\star}(\theta_j)$  minimizer of  $H(\cdot;\theta_j)$  such that  for all  $\epsilon >0$,
	\begin{equation}
	\label{eqn:convergence}
	\text{pr}\left[  \underset{j=1,\ldots,N}{\max}\,   \nu\left\{ \left(  A_{   \hat{w}^\star(\theta_j) }^S\backslash A_{   w^\star(\theta_j) }^S  \right)  \cup    \left(  A_{   w^\star(\theta_j) }^S \backslash  A_{   \hat{w}^\star(\theta_j) }^S\right) \right\}      \geq \epsilon\right]   \,\rightarrow\, 0,
	\end{equation}
	where  $\nu$  is the Lebesgue measure.
	Thus,  the acceptance regions for  the  grid points based on the   estimated spending function will be closed to those based on the  true spending function. Consequently, if a new data point  is collected then 
	the  confidence  set based on the estimated  spending  function and the  grid of points will be  equals to that  based on the same grid and the  true spending function, with probability approaching one   (see Corollary \ref{cor1} in Appendix   \ref{sec:ad_result}).  If  instead (\ref{eqn:violation})   holds for some $\theta_j$ or  $F_S(t;\theta_j)/\alpha$ is a minimizer of $H(\cdot;\theta_j)$,  then such  $\theta_j$ would have to be  excluded  from the statement in  (\ref{eqn:convergence})  and from that of Corollary \ref{cor1}. This is due to the discontinuity of $F_S^{-1}(;\theta_j)$ at  $F_S(t;\theta_j)$.    }
%the condition on $\Lambda$ is violated it will still be the case

On another note, we restrict ourselves to selective inference with a selection region of the form $S=\{y: |y| > t\}$ for $t > 0$.  However, because this result mostly hinges on convergence of $\hat m$ to $m$, it also holds for  asymmetric and one-sided selection regions. \response{In fact, for one sided regions the extra regularity condition on $q$  is not needed  for  Theorem  \ref{mytheorem}. Similarly, in such setting  (\ref{eqn:convergence})   and Corollary \ref{cor1}   hold without conditions on $q$. } Theorem \ref{mytheorem} also applies when constructing non-selective confidence sets, including the situation studied in \citet{YuHoff}.  A similar result is more difficult to show for conditional selection, as the selection-adjusted marginal in this case has the much less tractable form given in Eq.~\eqref{eq:conditional-marginal-est}.

We also note that the procedure comes in with a built-in form of robustness.  Even if the prior is not well estimated for a given sample, by construction the procedure will still give confidence sets that retain nominal coverage, as long as the spending function is not data-dependent.  In other words, prior misspecification can mitigate the efficiency of confidence sets, but not their coverage.

%%% Local Variables:
%%% mode: latex
%%% TeX-master: "../main"
%%% End:

%!TEX root = ../main.tex

\section{Results on simulated and real data}
\label{sec:results-simul-real}

\subsection{Case of well-specified prior}
\label{sec:case-well-specified}

We now investigate the performance of our saFAB procedure to the usual selective confidence set, i.e. the UMAU confidence sets developed by \cite{Reid2015}, at the $\alpha = 0.1$ level.  Specifically, we will check that nominal coverage is maintained, and also compare the efficiency (size) of the confidence sets. We consider saFAB confidence sets from three variants of constructing the spending function: (i) the ``oracle'' case where we know the parametric form and the hyperparameters of the prior, (ii) the ``parametric empirical Bayes'' (PEB) case where we assume a parametric form of the prior, and estimate hyperparameters via maximum marginal likelihood estimation using five-fold data-splitting, and (iii) the ``nonparametric empirical Bayes'' (NPEB) case as described in the \S\ref{sec:nonp-proc}, where estimate the prior $\pi(\theta)$ from the data via the predictive recursion algorithm of \cite{Newton02ona}, using five-fold data-splitting.

% \begin{enumerate}[(i)]
% \item The ``oracle'' case where we know the parametric form and the % hyperparameters of the prior.
% \item The ``parametric empirical Bayes'' (PEB) case where we assume a % parametric form of the prior, and estimate hyperparameters via % maximum marginal likelihood estimation using five-fold data-splitting.
% \item The ``nonparametric empirical Bayes'' (NPEB) case as described % in the \S\ref{sec:nonp-proc}, where estimate the prior % $\pi(\theta)$ from the data via the predictive recursion algorithm % of \cite{Newton02ona}, using five-fold data-splitting.
% \end{enumerate}

We sample $\theta_i$ from several considered distributions, and generate data $y_i$ from the Gaussian distribution centered on $\theta_i$ with unit variance.  The sampling variance is assumed known for each method of constructing confidence sets.

First we consider signals generated from the point mass-Gaussian mixture in Eq.~\eqref{eq:two-groups-normal} for the toy example in \S\ref{sec:toy-example} with the hyperparameters set to $p=0.2$ and $\tau^2 = 3$.  We simulate 2,000,000 total signals, structured in 1000 batches as follows.  For each batch we make $n = 2000$ draws from the mixture model for $\theta_i$ and generate $y_i$ from $N(\theta_i, 1)$.  For conditional selection, these $\theta_i$ are generated once and used for each batch, while for the joint selection case they are generated anew each time.  The selection rule for determining the $\theta_i$ for which to construct confidence sets is $S = \{y: |y| > 2\}$.  For these selected $\theta_i$ we construct 90\% confidence sets under the four considered methods, and then for each method we calculate the proportion of confidence sets which cover the true $\theta_i$, as well as the average size of the confidence sets.

Tables~\ref{tab:nonpar-joint} and \ref{tab:nonpar-cond} show the results for joint and conditional selection, respectively.  The coverage of each method hems closely to the nominal rate.  The oracle saFAB procedure, however, gives confidence sets that are on average approximately 10\% (joint selection) and 6\% (conditional selection) more efficient than the existing selective confidence set procedure (UMAU).  The efficiency of the parametric empirical Bayes saFAB procedure closely hems to that of the oracle.  Notably, the nonparametric empirical Bayes variant of saFAB also comes very close to the oracle, even though it makes no assumption of the form of the prior.

% \begin{align}\label{eq:two-groups-general}
%   \pi(\theta) = p \cdot \pi_1(\theta) + (1 - p) \cdot \delta_0(\theta)
% \end{align}

\begin{table}[t!]
\centering
\begin{tabular}{llll}
  \toprule
Method & Coverage        & Average size    & Relative average size \\ 
  \midrule
Oracle & 0.8990 (0.0095) & 3.3493 (0.0014) & 0.8956 (0.0004)       \\ 
  PEB  & 0.8990 (0.0095) & 3.3510 (0.0014) & 0.8960 (0.0004)       \\ 
  NPEB & 0.8995 (0.0095) & 3.3555 (0.0014) & 0.8972 (0.0004)       \\ 
  UMAU & 0.8993 (0.0095) & 3.7399 (0.0016) & 1.0000 (0.0004)       \\ 
   \bottomrule
\end{tabular}
\caption{
  Comparison of performance for the well specified prior case under joint selection.  We consider saFAB when the prior is assumed known (oracle), saFAB when the prior form is known and hyperparameters are estimated from the data (parametric empirical Bayes, PEB), saFAB when no prior form is assumed but is rather estimated from the data using predictive recusion (NPEB), and the existing selective confidence sets resulting from inverting unbiased tests (UMAU).
} \label{tab:nonpar-joint}
\end{table}

\begin{table}[t!]
\centering
\begin{tabular}{llll}
  \toprule
Method & Coverage        & Average size    & Relative average size \\ 
  \midrule
Oracle & 0.9005 (0.0095) & 3.5138 (0.0021) & 0.9385 (0.0006)       \\ 
  PEB  & 0.9004 (0.0095) & 3.5113 (0.0021) & 0.9378 (0.0006)       \\ 
  NPEB & 0.8999 (0.0095) & 3.5123 (0.0021) & 0.9381 (0.0005)       \\ 
  UMAU & 0.8998 (0.0095) & 3.7441 (0.0016) & 1.0000 (0.0004)       \\ 
   \bottomrule
\end{tabular}
\caption{Comparison of performance for the well specified prior case
  under conditional selection} \label{tab:nonpar-cond}
\end{table}

\subsection{Case of misspecified prior}

Now we compare the performance of the nonparametric procedure against the parametric procedure under a misspecified prior.  That is, for the parametric procedure we assume the same parametric form of a point mass-Gaussian mixture for the prior, but in reality the $\theta_i$ are drawn from a different distribution.  We study two different scenarios of misspecification: one where the non-zero $\theta_i$ come from a bimodal distribution with both modes separated from zero, that is
\begin{align*}
  \theta_i \sim (p/2) \mathcal{N}(-\mu, \tau^2) +
  (p/2) \mathcal{N}(\mu, \tau^2) +
  (1 - p) \delta_0 
  % \begin{cases}
  %   \mathcal{N}(-\mu, \tau^2) & \text{ w.p. } p/2 \\ 
  %   \mathcal{N}(\mu, \tau^2) & \text{ w.p. } p/2 \\
  %   \delta_0 & \text{ w.p. } 1-p
  % \end{cases}
\end{align*}
with $\tau^2=1/4$ and $p=0.2$, and one where they come from a skewed, unimodal nonzero-centered distribution, that is
\begin{align*}
  \theta_i \sim p \{\mu + \mathrm{Exponential}(\lambda)\} + (1 -
  p) \delta_0
  % \begin{cases}
  %   \mu + \mathrm{Exponential}(\lambda) &\text{ w.p. } p \\
  %   \delta_0 &\text{ w.p. } 1-p.
  % \end{cases}
\end{align*}
with $\mu = 1$ and $\lambda = 1$. 

% Now we compare the performance of the nonparametric procedure against
% the parametric procedure under a misspecified prior. We still assume
% the point mass-Gaussian mixture prior \eqref{eq:two-groups-normal} for
% the parametric procedure, but in actuality the $\theta_i$ will be
% drawn from a different distribution. 

In both cases, we simulate 1000 batches of $n = 2000$ pairs of $(\theta_i, y_i)$, and we retain the same selection region, $S = \{y: |y| > t \}$. For brevity, we only consider selection acting jointly on $(\theta_i, y_i)$.

Tables~\ref{tab:misspec-bimodal} and \ref{tab:misspec-skew} show the results.  Again, each method maintains nominal coverage.  That this is true for the parametric saFAB procedure shows that even under prior misspecification, saFAB gives correct coverage.  For both considered priors, the oracle saFAB gives confidence sets which are about 11--12\% more efficient than the UMAU procedure.  Interestingly, the parametric and nonparametric saFAB procedures show only minor differences in their efficiency gains, and come close to the oracle.  Whereas nonparametric slightly outperforms parametric in the case of the skewed alternative density, this result is reversed for the bimodal alternative density.  This latter fact is surprising given that the assumed prior in the parametric procedure is incorrect, and is likely explained by a high estimated variance for the unimodal alternative density.

% highlights the fact that misspecification retains the 

% \begin{figure}[t!]
%   \centering
%   \includegraphics[width=0.65\textwidth]{figures/lenplot.eps}
%   \caption{\label{fig:lenplot-skew} Comparing confidence set
%     efficiency, case of misspecified prior with skewed alternative
%     density}
% \end{figure}

\begin{table}[t!]
\centering
\begin{tabular}{llll}
  \toprule
Method & Coverage        & Average size    & Relative average size \\ 
  \midrule
Oracle & 0.9012 (0.0094) & 3.3941 (0.0007) & 0.8884 (0.0002)       \\ 
  PEB  & 0.9038 (0.0093) & 3.4148 (0.0007) & 0.8938 (0.0002)       \\ 
  NPEB & 0.9017 (0.0094) & 3.4335 (0.0008) & 0.8987 (0.0002)       \\ 
  UMAU & 0.8999 (0.0095) & 3.8204 (0.0009) & 1.0000 (0.0002)       \\ 
   \bottomrule
\end{tabular}
\caption{
  Comparison of performance when the prior is misspecified for the parametric empirical Bayes saFAB procedure.  The true density for $\theta$ is a mixture of a point mass at zero and a bimodal alternative density, and signals are selected under joint selection.
} \label{tab:misspec-bimodal}
\end{table}

\begin{table}[t!]
\centering
\begin{tabular}{llll}
  \toprule
Method & Coverage        & Average size    & Relative average size \\ 
  \midrule
Oracle & 0.8940 (0.0097) & 3.3149 (0.0011) & 0.8808 (0.0003)       \\ 
  PEB  & 0.8937 (0.0097) & 3.3941 (0.0011) & 0.9019 (0.0003)       \\ 
  NPEB & 0.9070 (0.0092) & 3.3672 (0.0019) & 0.8947 (0.0005)       \\ 
  UMAU & 0.9002 (0.0095) & 3.7635 (0.0014) & 1.0000 (0.0004)       \\ 
   \bottomrule
\end{tabular}
\caption{
  Comparison of performance when the prior is misspecified for the parametric empirical Bayes saFAB procedure.  The true density for $\theta$ is a mixture of a point mass at zero and a skewed alternative density, and signals are selected under joint selection.
} \label{tab:misspec-skew}
\end{table}

\subsection{Data-dependent thresholding}

So far we have so far presumed that the selection set $S$ is specified \emph{a priori}, so that the selection-adjusted sampling model is well-defined.  Now we investigate the use of data-adaptive selection sets\response{, where the selection set $S$ is now a random variable and so the truncated sampling model in Eq.~\eqref{eq:sa-likelihood} in not well-defined}.  We focus on the popular Benjamini-Hochberg (BH) procedure \citep{BH} for controlling the false discovery rate (FDR) to decide the selection rule.  To briefly summarize the BH procedure, suppose we have $m$ null hypotheses $H_{01}, \ldots, H_{0m}$ all with corresponding $p$-values $p_1, \ldots, p_m$, and let $p_{(k)}$ represent the $k$th smallest $p$-value with corresponding null hypothesis $H_{(k)}$.  Rejecting all null hypotheses $H_{(1)}, \ldots, H_{(k^\star)}$ with
\begin{align}\label{eq:BH}
  k^\star = \max \left \{k : {p_{(k)}}/{k} \leq {q^\star}/{m} \right \}  
\end{align}
bounds the FDR at $q^\star$, where the FDR is the expected proportion of rejected null hypotheses which are actually true. This is a data-dependent rule for selection because we must calculate the $p$-values from the data to decide the selection rule.
% Previously, \cite{FCR} developed confidence intervals for parameters
% selected by the BH procedure, but these intervals are needlessly
% conservative; 
\cite{Reid2015} characterize the BH procedure as an affine linear constraint for use in the selection-adjusted likelihood for constructing UMAU confidence intervals, but here we are interested in investigating the performance of confidence sets constructed by simply treating the BH procedure as a simple data-dependent thresholding problem.

% Reference Reid et al., and Benjamini 

We perform a simulation study to see if nominal coverage is still upheld using this approach.  We generate data in an identical way as done in \S\ref{sec:case-well-specified}.  We only consider the case of joint selection.  Each two-sided $p$-value for the point null hypothesis $H_{0i}: \theta_i = 0$ is then given by $p_i = 2 \Phi(-|y_i|)$, where $\Phi(\cdot)$ is the standard normal cumulative density function.  We use the BH rule in Eq.~(\ref{eq:BH}) to choose which $\theta_i$ to be considered for inference.  The selection set for constructing the selection-adjusted likelihood is now considered to be $S = \{y : |y| > |y_{(k^\star + 1)}| \}$, i.e. the threshold is determined by the largest $y_i$ in magnitude of those not chosen by the BH procedure.  We use $q^\star = 0.2$ as the target FDR for the decision rule, and again construct 90\% confidence sets for each method, performing 1000 simulations.

The results are presented in Table~\ref{tab:bh-joint}. We appear to recover nominal coverage for each method. This suggests that using the BH procedure as a data-dependent selection rule will still give valid confidence sets for the saFAB procedure.

\begin{table}[t!]
\centering
\begin{tabular}{lccc}
  \toprule
Method & Average coverage & Average size    & Relative average size \\ 
  \midrule
Oracle & 0.9028 (0.0094)  & 3.7088 (0.0019) & 0.9158 (0.0005)       \\ % \hline
  PEB  & 0.9033 (0.0093)  & 3.7072 (0.0019) & 0.9154 (0.0005)       \\ 
  NPEB & 0.9032 (0.0093)  & 3.7084 (0.0022) & 0.9157 (0.0005)       \\ 
  UMAU & 0.9088 (0.0091)  & 4.0500 (0.0014) & 1.0000 (0.0003)       \\
  \bottomrule
\end{tabular}
\caption{
  Comparison of performance when using a data-dependent threshold, via the Benjamini-Hochberg procedure, for the selection set, and signals are selected under joint selection.
} \label{tab:bh-joint}
\end{table}

%%% Local Variables:
%%% mode: latex
%%% TeX-master: "../main"
%%% End:

%!TEX root = ../main.tex

\subsection{Analysis of neural synchrony data}
\label{sec:analys-neur-synchr}

%\begin{figure}[ht!]
%  \centering
%  \includegraphics[width=0.7\textwidth]{figures/cortex-spendfun.eps}
%  \caption{\label{fig:cortex-spendfun} Spending function calculated
%    for neural synchrony data. }
%\end{figure}

\begin{figure}[ht!]
  \centering
  \includegraphics[width=0.6\textwidth]{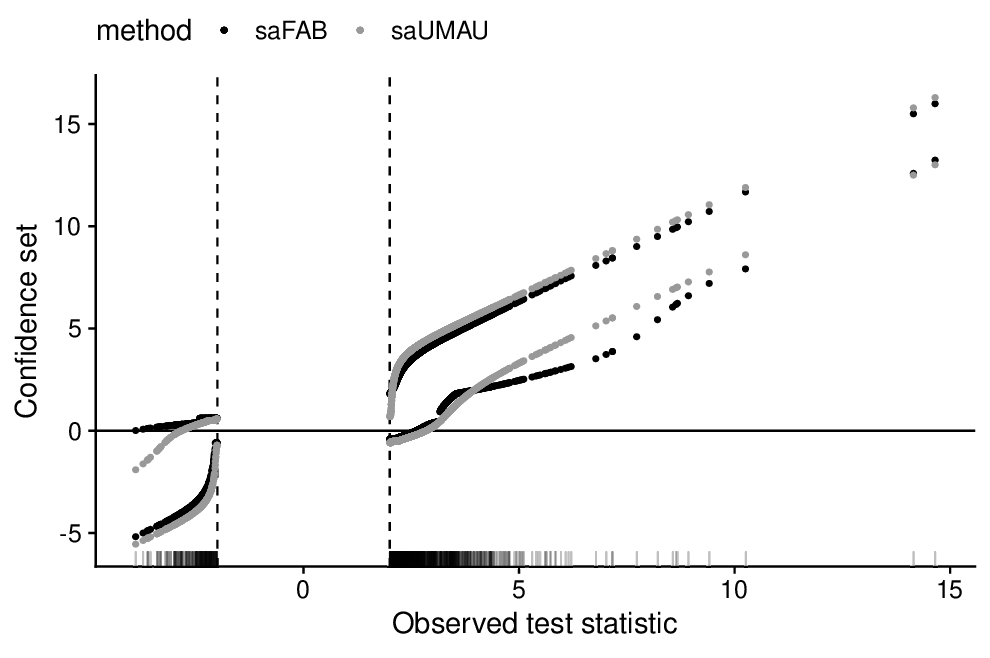}
  \caption{\label{fig:cotex-intervals}
    Comparison of confidence intervals from saFAB and UMAU.  Our saFAB procedure gives confidence intervals that are 11.3\% shorter on average, with 85\% of constructed intervals being shorter.
  }
\end{figure}

Finally, we apply the nonparametric saFAB procedure to a real dataset to show how, with minimal assumptions about the prior, we may still construct confidence sets which have significant gains in efficiency over the UMAU confidence sets.  To do so, we analyze the neural synchrony data published in \cite{Smith12591} and \cite{Kelly2010}, and re-analyzed by \cite{FDRreg}.  The goal of this application is to identify fine-time-scale neural interactions (``synchrony'') among many neurons recorded simultaneously by a multi-electrode array.  The experiment from which the data are drawn produced thousands of pairwise test statistics, each representing the magnitude of interaction between a single pair of neurons.

\cite{Kelly2010} provide full details of the data and the experiment. For our purposes, the relevant fact is that the data for each neuron pair can be assumed to take the form $z_i \sim N(\theta_i, 1)$ (we use $i$ to index pairs, which can be thought of as edges in a network).  Here $\theta_i$ can be interpreted as a log rate ratio: that is, $e^{\theta_i}$ represents how much more often, in multiplicative terms, the two neurons in pair $i$ fire together, compared to the rate one would expect if they were firing independently.  Thus if $\theta_i \approx 0$, the two neurons are plausibly independent, while if $\theta_i$ is substantially larger than zero, they exhibit an interesting pattern of fine-time-scale synchronous firing.  The case $\theta_i < 0$ is less well understood scientifically, but potentially interesting as well.

In our analysis, we assume no parametric form of the prior for $\theta$, preferring instead to use the nonparametric saFAB procedure outlined in \S\ref{sec:nonp-proc} to estimate the prior via predictive recursion.  We use $S = \{y : |y| > 2 \}$ as a very liberal selection region.  Here selection acts jointly on $\theta_i$ and $y_i$, since we will form confidence sets only for those $\theta_i$ that meeting an initial screen of significance.
% 
% Figure~\ref{fig:cortex-spendfun} shows the constructed spending function, and 
Figure~\ref{fig:cotex-intervals} shows the saFAB confidence sets as compared to the selection-adjusted UMAU confidence sets.  The asymmetry in these intervals reflects that fact that most signals in the data set corresponding to presumed cases where $\theta_i > 0$ (synchrony enhancement), rather than $\theta_i < 0$ (synchrony suppression).  Indeed, there are clear scientific reasons to suspect than many neuron pairs will have $\theta_i > 0$, but the case $\theta_i < 0$ would be unusual.  Our analysis detects this fact without having to assume it, and adapts to it via the choice of the spending function.

For the large majority selected test statistics, about 85\%, our saFAB method produces a shorter interval than the UMAU method.  The saFAB intervals are considerably more efficient; the average width of the saFAB intervals is 3.38, about 11.3\% smaller than the average width of UMAU intervals, which is 3.81.  While the saFAB intervals are wider for some of the larger test statistics, these observations are relatively few.  Indeed, because the marginal distribution of selected test statistics is tightly concentrated near the boundary of the selection region, these are the intervals which receive the efficiency gains of our procedure by design.  Having tighter intervals for these observations is important since there is a greater chance of them being false positives.

We emphasize that our analysis complements, rather than competes with, the analyses in \citet{Kelly2010} and \cite{FDRreg}.  In those papers, the goal was to discover interesting pairs of neurons, i.e.~to test which pairs have $\theta_i \neq 0$.  Our analysis takes such a test as a starting point.  Using the techniques developed here, we are able conduct valid frequentist inference for the discovered $\theta_i$'s, while exploiting their probabilistic structure via a prior, and simultaneously controlling for post-selection inference.  Because $\theta_i$ has a useful neurophysiological interpretation as a log relative rate, quantifying uncertainty about its magnitude can in this manner can add substantially to the analyses conducted by previous authors.

%%% Local Variables:
%%% mode: latex
%%% TeX-master: "../main"
%%% End:

%!TEX root = ../main.tex

\section{Discussion}
\label{sec:discussion}

The central argument of this paper has been that the use of a prior can play a decisive and favorable role in constructing optimal selection-adjusted confidence sets. Here we identify three open questions.  First, an interesting future line of work would be to undertake a formal investigation of the performance of the saFAB method under data-dependent thresholding, such as the Benjamini--Hochberg procedure.  Our simulation-based results suggest that the performance is excellent, while theoretical investigations of the UMAU method show that coverage does indeed hold because the selection rule satisfies a key polyhedral condition \citep[e.g.][]{lee2016}.  But more theoretical work is necessary to merge these lines of reasoning.  \response{ Second, it may be of interest to create intervals which have minimum expected size with respect to the marginal distribution of selected signals \citep[i.e. $\pi_S(\theta)$ in the notation of][]{saBayes} rather than with respect to the entire distribution of signals, i.e. $\pi(\theta)$.  We leave avenue this to future work. } Finally, there is much research left to be done in this direction of merging Bayesian and frequentist thinking in other areas of inference where selection takes place.  Current ongoing work extends our methods to regression, spatial hotspot detection, and subgroup identification in causal inference. Preliminary results show promising improvements in these areas.

\appendix

% \appendixone

\section{Predictive recursion}
\label{sec:predictive-recursion}

For the nonparametric saFAB procedure, we assume that the data arise from the model
\begin{align*}
  \begin{split}
    y_i & \sim f(y_i; \theta_i), \quad
    \theta_i \sim \pi(\theta) 
  \end{split}
\end{align*}
% By marginalizing out $\theta$ we may reformulate the model as
% \begin{align*}
%   \begin{split}
%     y_i    & \sim p \cdot f_1(y_i) + (1 - p) \cdot f_0(y_i) \\
%     f_0(z) & \sim \mathcal{N}(y; 0, \sigma^2)               \\
%     f_1(z) & \sim
%     \int_\mathbb{R} \mathcal{N}(y_i; \theta, \sigma^2) \pi(\theta) \mathrm{d}\theta.
%   \end{split}
% \end{align*}
where the sampling density $f(y; \theta)$ is known, and the prior $\pi(\theta)$ is to be estimated.  To do so, we use predictive recursion method of \cite{Newton02ona} to estimate the mixing density $\pi(\theta)$ from the observations $y_1, \ldots, y_n$.

We begin with an intial guess $\pi^{[0]}$ and a sequence of weights $\gamma^{[i]} \in (0, 1)$. For $i=1,\ldots,n$, we recursively compute the update
\begin{align}
  m^{[i - 1]}(y_i) &= \int_\mathbb{R} f(y_i; u) \pi(\mathrm{d} u) \label{eq:mmm}\\
  \pi^{[i]}(\mathrm{d}u) &=
  (1 - \gamma^{[i]}) \pi^{[i-1]}(\mathrm{d}u) +
  \gamma^{[i]}  \frac{f(y_i; \theta) \pi^{[i-1]}(\mathrm{d}u)}{m^{[i-1]}(y_i)}.
  \nonumber
\end{align}
% \begin{align}
%   m^{[i - 1]}(y_i) &= \int_\mathbb{R} \mathcal{N}(y_i; u, \sigma^2)
%                      \Psi^{[i - 1]}(\mathrm{d}u)  \label{mmm}\\
%   \Psi^{[i]}(\mathrm{d}u) &= (1 - \gamma^{[i]})
%                             \Psi^{[i-1]}(\mathrm{d}u) + \gamma^{[i]}
%                             \cdot 
%                             \left\{
%                             \frac{
%                             \mathcal{N}(y_i; u, \sigma^2)
%                             \Psi^{[i-1]}(\mathrm{d}u)
%                             }
%                             {m^{[i-1]}(y_i)}
%                             \right\}  \nonumber
% \end{align}

Algorithm \ref{al:pr} details our implementation. We sweep through the data 10 times, each time randomizing the sweep order over the data. In practice, the prior $\pi(\theta)$ is computed on a grid, and the integral in Eq.~\eqref{eq:mmm} is computed using the trapezoid rule. \cite{Tokdaretal} give conditions on the weights $\gamma^{[i]}$ to lead to almost-sure weak convergence of the PR estimate to the true mixing distribution. In the case that the mixture model is misspecified, they show that the PR estimate converges in total variation to the mixing density that minimizes the Kullback-Leibler divergence to the truth. Specifically, the conditions for convergence are satisfied by $\gamma^{[i]} = (i + 1)^{-a}$, and we use the default value $a = 0.67$ recommended by \cite{Tokdaretal}.
% More details on 
% \cite{Newton02ona}, \cite{MartinTokdar}, and \cite{Tokdaretal}.

\begin{algorithm}
  \SetKwInOut{Input}{Input}
  \SetKwInOut{Output}{Output}
  
  \Input{Data $y_1, \ldots, y_n$; sampling model
    $y_i \sim f(y_i; \theta_i)$; intial guess
    $\pi^{[0]}(\theta)$. }
  
  \For{$i = 1,\ldots,n$}{
    \begin{align*}
      % m_0^{[i]} &= \pi_0^{[i - 1]} \cdot \mathcal{N}(y_i; 0, \sigma^2) \\
      g^{[i]}(\theta) &= f(y_i; \theta)
                          {\pi}_1^{[i - 1]}(\theta) \quad
                          \text{(discrete grid)} \\
      m^{[i]} &= \int_\mathbb{R}g^{[i]}(\theta) \mathrm{d}\theta \quad
                  \text{(trapezoid rule)} \\
      % \pi_0^{[i]} &= (1 - \gamma^{[i]}) \cdot \pi_0^{[i - 1]} +
      %               \gamma^{[i]} \cdot 
      %               \left(
      %               \frac{m_0^{[i]}}{m_0^{[i]} + m_1^{[i]}}
      %               \right) \\
      \pi^{[i]}(\theta) &= (1 - \gamma^{[i]}) \cdot \pi_1^{[i - 1]}(\theta) +
                    \gamma^{[i]} \cdot 
                    \left(
                    \frac{f(y_i; \theta) \pi^{[i-1]}(\theta)}{m^{[i]}}
                    \right)
    \end{align*}
  }

  \Output{Estimate $\hat \pi(\theta) = \pi^{[n]}(\theta)$. } \BlankLine
  \caption{Predictive recursion algorithm for estimating the prior,
    which is then used for estimating the optimal spending function as
    outlined in \S\ref{sec:nonp-proc}.}\label{al:pr}
\end{algorithm}\DecMargin{1em}

% \begin{algorithm}
%   \SetKwInOut{Input}{Input}
%   \SetKwInOut{Output}{Output}
  
%   \Input{Data $y_1, \ldots, y_n$; null model
%     $\mathcal{N}(0, \sigma^2)$; intial guess
%     $\Psi^{[0]} = \tilde{\pi}_1^{[0]}(\theta) + \pi_0^{[0]} \delta_0$
%     with continuous subdensity $\tilde{\pi}_1^{[0]}(\theta)$ and a
%     Dirac measure at zero of mass $\pi_0^{[0]}$}
  
%   \For{$i = 1,\ldots,n$}{
%     \begin{align*}
%       m_0^{[i]} &= \pi_0^{[i - 1]} \cdot \mathcal{N}(y_i; 0, \sigma^2) \\
%       f_1^{[i]}(\theta) &= \mathcal{N}(y_i; \theta, \sigma^2)
%                           \tilde{\pi}_1^{[i - 1]}(\theta) \quad
%                           \text{(discrete grid)} \\
%       m_1^{[i]} &= \int_\mathbb{R}f_1^{[i]}(\theta) \mathrm{d}\theta \quad
%                   \text{(trapezoid rule)} \\
%       \pi_0^{[i]} &= (1 - \gamma^{[i]}) \cdot \pi_0^{[i - 1]} +
%                     \gamma^{[i]} \cdot 
%                     \left(
%                     \frac{m_0^{[i]}}{m_0^{[i]} + m_1^{[i]}}
%                     \right) \\
%       \pi_1^{[i]}(\theta) &= (1 - \gamma^{[i]}) \cdot \pi_1^{[i - 1]} +
%                     \gamma^{[i]} \cdot 
%                     \left(
%                     \frac{f_1^{[i]}(\theta)}{m_0^{[i]} + m_1^{[i]}}
%                     \right)
%     \end{align*}
%   }

%   \Output{Estimates $p = 1 - \pi_0^{[n]}$ and
%     $\pi_1(\theta) = \tilde{\pi}_1^{[n]}(\theta) / p$} \BlankLine
%   \caption{Predictive recursion}\label{al:pr}
% \end{algorithm}\DecMargin{1em}

%%% Local Variables:
%%% mode: latex
%%% TeX-master: "../main"
%%% End:

\section{Assumptions for main result}
\label{sec:asumptions}
    \response{Here we state the assumptions  from \cite{Tokdaretal}  that are required for  Theorem \ref{mytheorem}.\\
A1. The weights  $\{\gamma^{[i]}\}_{i=1}^{\infty}$ in Algorithm \ref{al:pr} are chosen to satisfy $\sum_{i=1}^{\infty} \gamma^{[i]} = \infty  $ and $\sum_{i=1}^{\infty} (\gamma^{[i]})^2 < \infty  $.\\
A2. The map  $F \rightarrow \int   f(y;\theta) F(d\theta)$  is injective.\\
A3. For  each  $y$,  the map $\theta \rightarrow   f(y;\theta)$ is  bounded and  continuous.\\
A4. For  all  $\epsilon>0$,   and     $\mathcal{X}_0 $  compact set,    there  exists  a compact  set  $\Theta_0$  such that   $\int_{ \mathcal{X}_0}     f(y;\theta)  dy  <\epsilon   $ for all $\theta \notin  \Theta_0$. \\
A5.  There  exists  a constant  $B<\infty$  such that  for  all $\theta_1, \theta_2$  and  $\theta_3$  we have that
\[
\int    \left\{  \frac{   f(y;\theta_1)   }{f(y;\theta_2)}  \right\}^2   f(y;\theta_3) dy  \,<\,B.\\  
\]
}

\section{Proof of main result}
\label{sec:proof-main-result}

 Throughout we drop the dependence  on $\theta$  and simply refer to $H(\cdot,\theta)$  as $H$.  
To  prove Theorem~\ref{mytheorem},  we first notice  that
\[
\underset{y \in R }{\sup} \,\,\vert M_S(y) - \hat M_S(y) \vert  \leq   \int \vert   m_S(y) - \hat m_S(y)  \vert dy \rightarrow^{a.s} 0,
\]
by Theorem 2 from \cite{Tokdaretal}.  Hence,  $\hat{H} \rightarrow H$  uniformly, almost surely.

Next, we observe that  the function $M_S$ is continuous everywhere as by Fubini's theorem
\begin{align*} M_S(y) = \int_{-\infty}^{y} m_S(u)\dee u
= \int_{-\infty}^y \int f_S(u;\theta) \pi(\dee\theta)\dee u =
\int \int_{-\infty}^y f_S(u;\theta) \dee u \pi(\dee \theta) = \int
F_S(y; \theta) \pi(\dee \theta).
\end{align*}
Furthermore, the function $F_S^{-1}$ is continuous everywhere except at $q = F_S(-t; \theta) = F_S(t; \theta)$, a point where $F_S^{-1}$ is left continuous.  Therefore, $H$ is continuous everywhere except perhaps in a set $\Lambda$ containing at most two different points. Additionally, $H$ is left continuous everywhere.

With the above in mind, the proof of (\ref{eqn:con1}) uses Wald's argument, in the spirit of Theorem 5.14 from \cite{vaart_1998}, or the original paper  \cite{wald1949}.  We start by taking $w_0 \in (0,1)$ minimizer of $H$. We also set
\[
  \mathcal{T} := [0,1] \backslash \left(  \Omega_0 \cup \Lambda  \right),
\]
where  $\Lambda $ is the set of discontinuities of $H$.  Then  $\Lambda \subset \{   (q+\alpha-1)/\alpha ,q/\alpha \}$.  If $w =  (q+\alpha-1)/\alpha \in [0,1)$, we observe that
\[
\begin{array}{lll}
   \underset{ \tilde{w}  \to w^+   }{\lim}  \,H(\tilde{w}) &=& M_S\left[    \underset{ \tilde{q}  \to q^+   }{\lim} F_S^{-1}(    \tilde{q}    )  \right ] -  M_S[F_S^{-1}(    q  +\alpha-1  )   ]  \\
   &>& M_S[F_S^{-1}(    q    )   ] -  M_S[F_S^{-1}(    q  +\alpha-1  )   ]  \\
    & =&  H(w)\\
    %\underset{\tilde{q}  \to q^{-}  }{\lim}\,  \left\{M_S[F_S^{-1}(    \tilde{q}    )   ] -  M_S[F_S^{-1}( \tilde{q}+\alpha-1  )   ] \right\} \\ 
 & \geq& \underset{ \tilde{w} \in [0,1] }{\inf} H(\tilde{w} ). 
\end{array}
\]
Therefore,  by our assumption on $q$  (when $q\leq  \alpha$), we have that   
\[
  \underset{ w \in \Lambda  \backslash\Omega_0  }{\min}   \,\underset{ \tilde{w}  \to w^+   }{\lim}  \,H(\tilde{w})   -    \underset{ w \in [0,1] }{\inf} H(w) >0.
\]
Hence, for  $\epsilon>0$ is small enough it holds that 
\begin{equation}
	\label{eqn:cond3}
  \underset{ w \in \Lambda  \backslash\Omega_0  }{\min}  \,\underset{ \tilde{w}  \to w^+   }{\lim}  \,H(\tilde{w})    -    \underset{ w \in [0,1] }{\inf} H(w)  >  2\epsilon.
\end{equation}
% , which exists since $[0,1]$ is
% compact and $L$ is continuous.  
Next, let $w \in  \mathcal{T}$ be fixed, and let $U_l $ be a decreasing sequence of intervals around $w$, with $\text{diameter}(U_l)$ converging to zero. Let
\begin{align*} H(U) = \underset{\tilde{w} \in U }{\inf }
\,H(\tilde{w}),
\end{align*}
for $U \subset [0,1]$.  Notice that $H(U_l) \geq H(w_0)$ for all $l$. Suppose that $ \lim_{l \to \infty } H(U_l)    =  H(w_0)$.  Then there exists a sequence $\ \{w_l\}$ with $w_l \in U_l$ such that $H(w_l) \rightarrow H(w_0)$. Hence, by the Bolzano--Weierstrass theorem, we have that for a subsequence of $\{w_l\}$, say $\{w_{l_k}\}$ it holds that $w_{l_k} \rightarrow w^{\prime}$ for some $w^{\prime } \in [0,1]$. However, since $U_{l_k}$ converges to $w$, it must be the case that $w^{\prime} =w$. Hence,
\begin{equation}
\label{eqn:contradiction}
 H(w) \, = \, H(w^{\prime }) \, = \,
\underset{k \rightarrow \infty}{\lim }\, H(w_{l_k}) \,=\, H(w_0),
\end{equation}
which contradicts  $w \in  \mathcal{T}$. Therefore, for all $w  \in  \mathcal{T}$ we have that $H(U_w) > H(w_0)$ for some small enough neighborhood $U_w$ of $w$.

If  $\Lambda   \backslash \Omega_0 \neq \emptyset$,  let   $w \in \Lambda \backslash \Omega_0$  and let $U_l $ be a decreasing sequence of intervals around $w$, with $\text{diameter}(U_l)$ converging to zero.   Suppose that $ \lim_{l \to \infty } H(U_l)    =  H(w_0)$.  Then there exists a sequence $\ \{w_l\}$ with $w_l \in U_l$ such that $H(w_l) \rightarrow H(w_0)$. Hence, by the Bolzano--Weierstrass theorem, we have that  for a subsequence of $\{w_l\}$, say $\{w_{l_k}\}$ it holds that $w_{l_k} \rightarrow w$,  and either  $w_{l_k} \geq  w$  for all $k$  or  $w_{l_k} \leq  w$  for all $k$. If   $w_{l_k} \geq  w$  for all $k$  then 
\[
   \underset{ \tilde{w} \to  w^+  }{\lim }H(w)  =  \underset{ k\to  \infty   }{\lim }H(w_{l_k}) =  H(w_0) , 
\]
which  contradicts (\ref{eqn:cond3}). On the other hand if, $w_{l_k} \leq  w$  for all $k$,  then by the left continuity of  $H$ we arrive at  (\ref{eqn:contradiction})  which once again  contradicts (\ref{eqn:cond3}).  Therefore, for all $w  \notin \Omega_0$ we have that $H(U_w) > H(w_0)$ for some small enough neighborhood $U_w$ of $w$.

% 
% -------------------------------------------------------------------------

\iffalse Then notice that
% 
\begin{align*} H^{\prime}(w) = \alpha\left[ \frac{ m(
F^{-1}( 1-\alpha + \alpha w ) ) }{f( F^{-1}( 1-\alpha + \alpha w ) )}
\,-\, \frac{ m( F^{-1}( \alpha w ) ) }{f( F^{-1}( \alpha w )
)}\right],
\end{align*}
% 
which is a continuous function and its well defined in $(0,1)$.  If $w\in (0,1)$, then
\[
  \begin{array}{lll} \vert H(U_l) - H(w) \vert & \leq &
\underset{ \tilde{w} \in U_l }{\sup}\, \vert H(\tilde{w}) - H(w)
\vert\\ & \leq& \mathrm{diameter}(U_l) \, \,\, \underset{ \tilde{w}
\in \overline{U}_1 }{\sup} \, H^{\prime}( \tilde{w} )\\ & \underset{l
\small\rightarrow \infty}{\rightarrow} & 0, \,\,\,\,\,\,
  \end{array}
\]
	
where the second inequality holds by the mean value theorem, and the last since $H^{\prime }(\cdot)$ is continuous in $\overline{U}_1$ (the closure of $U_1$). Therefore, for all $w \notin \Omega_0$ we have that $H(U_w) > H(w_0)$ for some small enough neighborhood $U_w$ of $w$. \fi

% -------------------------------------------------------------------------

Now, notice that the set $B$ is compact, where
\begin{align*} B := \left\{w \in [0,1] \,:\,\,\,\,\,
\underset{\tilde{w} \in \Omega_0   }{\inf}\, | w - \tilde{w} | \geq
\epsilon \right\}.
\end{align*}
Clearly, $B$ can be covered by the intervals $\{ U_w \,:\, w\in B \}$.  Hence, there exists $U_{w^{(1)}},\ldots, U_{w^{(p)} }$ that cover $B$. Therefore,
\begin{align} \label{eqn:ineq}
  \underset{ w \in B }{\inf}\, \hat{H}(w) \,\geq \,
\underset{ w \in \cup_{j=1}^p U_{ w^{(j)}} }{\inf}\,
\hat{H}(w)\,\rightarrow_{ \text{a.s} } \underset{ w \in \cup_{j=1}^p U_{
w^{(j)}} }{\inf}\, H(w) \,>\, H(w_0),
\end{align}
where the limit follows from the, almost sure, uniform convergence of $\hat{H}$ to $H$. However, if $\hat{w}^\star \in B \cap \hat{\Omega}_n$, then
\begin{align*}
  \underset{ w \in B \cap \hat{\Omega}_n}{\inf}\, \hat{H}(w) \, = \, \underset{ w \in B }{\inf}\, \hat{H}(w)\,=\,
  \underset{ w \in [0,1] }{\inf}\, \hat{H}(w) \,\leq \, \underset{ w \in
  [0,1] }{\inf}\, H(w) + o_P(1).
\end{align*}
Therefore,
\begin{align*}
  \left\{ \hat{w}^\star \in B \cap \hat{\Omega}_n
\right\}\,\subset\,\left\{ \underset{ w \in B }{\inf}\, \hat{H}(w) \,\leq
\, H(w_0) + o_P(1) \right\},
\end{align*}
but, according to (\ref{eqn:ineq}), the event on the right has probability that converges to zero as $n \rightarrow 0$.  This proves (\ref{eqn:con1}).   To prove (\ref{eqn:con2}), notice that
\begin{align*}
  0 \leq H(\hat{w}^\star) - H(w^\star) \leq 
  \hat{H}(\hat{w}^\star) - \hat{H}(w^\star) + 2\underset{\tilde{w} \in
    [0,1] }{\sup} \vert H(\tilde{w}) - \hat{H}( \tilde{w} ) \vert
  \leq  2\underset{\tilde{w} \in [0,1] }{\sup} \vert
  H(\tilde{w}) - \hat{H}( \tilde{w} ) \vert \rightarrow_{\text{a.s}}
  0,
\end{align*}
and so, $ H(\hat{w}^\star) \rightarrow H(w^\star) $ almost surely, where $w^\star \in \Omega_0$.

\section{Additional result}
\label{sec:ad_result}
\response{\begin{corollary}
	\label{cor1}  Let  $z$ be a random variable with  probability density function  bounded by above.   Let  $\theta_1,\ldots,\theta_N$  be fixed points  in the real line. Suppose that all the assumptions in Theorem \ref{mytheorem} hold for  each  $\theta_j$   and the set $\{0,F_S(t;\theta_j)/\alpha,1\}$ contains no  minimizer of $H(\cdot;\theta_j)$ for  all  $j =1,\ldots, N$. Then for  each $j$    and  $\hat{w}^\star(\theta_j)$ minimizer of $\hat H(;\theta_j)$   there exists a  $w^{\star}(\theta_j)$  minimizer of  $H(\cdot;\theta_j)$  such that 
	\[
	 \underset{n\to \infty }{\lim}\,    \mathrm{pr}\left(    \left\{  \theta_j \,:\,   z \in A_{  w^\star(\theta_j)  }^S,\,\,j \in  \{1,\ldots,N  \}  \right\} =   \left\{  \theta_j \,:\,   z \in A_{  \hat{w}^\star(\theta_j)  }^S,\,\,j \in  \{1,\ldots,N  \}  \right\}  \right)  \,=\,1.
	\]
 
\end{corollary}
%\{  \theta_j \,:\,   z \in A_[  w^*(\theta_j)  }^S,\,\,j \in  \{1,\ldots,N  \}   =  \{  \theta_j \,:\,   z \in A_[  w^*(\theta_j)  }^S,\,\,j \in  \{1,\ldots,N  \} 
\begin{proof}
	First, it is immediate  from Theorem  \ref{mytheorem}  that  (\ref{eqn:convergence}) holds.  Next, let  $g$  be the probability density function of  $g$.  Then
  \[
  	\begin{array}{l}
 	 \underset{n\to \infty }{\lim}\,    \text{pr}\left(    \left\{  \theta_j \,:\,   z \in A_{  w^\star(\theta_j)  }^S,\,\,j \in  \{1,\ldots,N  \}  \right\} =   \left\{  \theta_j \,:\,   z \in A_{  \hat{w}^\star(\theta_j)  }^S,\,\,j \in  \{1,\ldots,N  \}  \right\}  \right)  \\
\displaystyle 	 =\,\,  \underset{n\to \infty }{\lim}\,  \int  \text{pr}\left(    \left\{  \theta_j \,:\,   z \in A_{  w^\star(\theta_j)  }^S,\,\,j \in  \{1,\ldots,N  \}  \right\} =   \left\{  \theta_j \,:\,   z \in A_{  \hat{w}^\star(\theta_j)  }^S,\,\,j \in  \{1,\ldots,N  \}  \right\}  \bigg| z\right) g(z) dz \\
\displaystyle =\,\, \int\underset{n\to \infty }{\lim}\,  \text{pr}\left(    \left\{  \theta_j \,:\,   z \in A_{  w^\star(\theta_j)  }^S,\,\,j \in  \{1,\ldots,N  \}  \right\} =   \left\{  \theta_j \,:\,   z \in A_{  \hat{w}^\star(\theta_j)  }^S,\,\,j \in  \{1,\ldots,N  \}  \right\}  \bigg| z\right) g(z) dz \\
=\,\,1,
  \end{array} 
  \]
  where the second and third inequalities follow from the dominated convergence theorem and  (\ref{eqn:convergence}).
\end{proof}}

% Suppose that the latter event
% holds.  Then by Bolzano--Weierstrass theorem, there exists a
% subsequence $\{\hat w_{k}^\star\}$ of $\{\hat{w}^\star\}$ such that $\hat w_{k}
% \rightarrow w_0$ for some $w_0 \in \Omega_0$. The claim follows from
% the continuity of $L$.

%%% Local Variables:
%%% mode: latex
%%% TeX-master: "../main"
%%% End:

% \input{appendix/construct-spending-function}
% \input{appendix/one-sided-selection}

% -------------------------------------------------------------------------
% Bibliography

\bibliography{main}
\end{document}